\documentclass[runningheads]{llncs}
\usepackage[T1]{fontenc}
\usepackage{graphicx}
\usepackage{marvosym}
\usepackage[colorlinks=true,linkcolor=blue, citecolor=blue, urlcolor=black]{hyperref}
\usepackage{cite}
\usepackage{cleveref}
\usepackage{xcolor}
\usepackage{listings}
\usepackage{longtable}
\usepackage{booktabs}
\newcommand\code[1]{{\tt\small #1}}
\definecolor{dkgreen}{rgb}{0,0.3,0}
\definecolor{cmtgreen}{rgb}{0,0.6,0}
\definecolor{ltblue}{rgb}{0,0.4,0.4}
\definecolor{dkviolet}{rgb}{0.3,0,0.5}
\definecolor{dkblue}{rgb}{0,0.2,0.2}
\lstdefinelanguage{Coq}{ 
    mathescape=true,
    texcl=false, 
    escapeinside={(@}{@)},
    morekeywords=[1]{Section, Module, End, Require, Import, Export,
        Variable, Variables, Parameter, Parameters, Axiom, Hypothesis,
        Hypotheses, Notation, Local, Tactic, Reserved, Scope, Open, Close,
        Bind, Delimit, Definition, Let, Ltac, Fixpoint, CoFixpoint, Add,
        Morphism, Relation, Implicit, Arguments, Unset, Contextual,
        Strict, Prenex, Implicits, Inductive, CoInductive, Record,
        Structure, Canonical, Coercion, Context, Class, Global, Instance,
        Program, Infix, Theorem, Lemma, Corollary, Proposition, Fact,
        Remark, Example, Proof, Goal, Save, Qed, Defined, Hint, Resolve,
        Rewrite, View, Search, Show, Print, Printing, All, Eval, Check,
        Projections, inside, outside, Def},
    morekeywords=[2]{forall, exists, exists2, fun, fix, cofix, struct,
        match, with, end, as, in, return, let, if, is, then, else, for, of,
        nosimpl, when},
    morekeywords=[3]{Type, Prop, Set, true, false, option},
    morekeywords=[4]{pose, set, move, case, elim, apply, clear, hnf,
        intro, intros, generalize, rename, pattern, after, destruct,
        induction, using, refine, inversion, injection, rewrite, congr,
        unlock, compute, ring, field, fourier, replace, fold, unfold,
        change, cutrewrite, simpl, have, suff, wlog, suffices, without,
        loss, nat_norm, assert, cut, trivial, revert, bool_congr, nat_congr,
        symmetry, transitivity, auto, split, left, right, autorewrite},
    morekeywords=[5]{by, done, exact, reflexivity, tauto, romega, omega,
        assumption, solve, contradiction, discriminate},
    morekeywords=[6]{do, last, first, try, idtac, repeat},
    morecomment=[s]{(*}{*)},
    showstringspaces=false,
    morestring=[b]",
    morestring=[d]’,
    tabsize=3,
    extendedchars=false,
    sensitive=true,
    breaklines=false,
    basicstyle=\fontsize{9.5pt}{11.4pt}\selectfont\ttfamily,
    captionpos=b,
    columns=[l]flexible,
    identifierstyle={\ttfamily\color{black}},
    keywordstyle=[1]{\bfseries\ttfamily\color{dkviolet}},
    keywordstyle=[2]{\bfseries\ttfamily\color{dkgreen}},
    keywordstyle=[3]{\bfseries\ttfamily\color{ltblue}},
    keywordstyle=[4]{\bfseries\ttfamily\color{dkblue}},
    keywordstyle=[5]{\bfseries\ttfamily\color{dkred}},
    stringstyle=\ttfamily,
    commentstyle={\bfseries\ttfamily\color{cmtgreen}},
    literate=
    {\\forall}{{\color{dkgreen}{$\forall\;$}}}1
    {\\exists}{{$\exists\;$}}1
    {<-}{{$\leftarrow\;$}}1
    {=>}{{$\Rightarrow\;$}}1
    {==}{{\code{==}\;}}1
    {==>}{{\code{==>}\;}}1
    {->}{{$\rightarrow\;$}}1
    {<->}{{$\leftrightarrow\;$}}1
    {<==}{{$\leq\;$}}1
    {\#}{{$^\star$}}1 
    {\\o}{{$\circ\;$}}1 
    {\@}{{$\cdot$}}1 
    {\/\\}{{$\wedge\;$}}1
    {\\\/}{{$\vee\;$}}1
    {++}{{\code{++}}}1
    {~}{{$\sim$}}1
    {\@\@}{{$@$}}1
    {\\mapsto}{{$\mapsto\;$}}1
    {\\hline}{{\rule{\linewidth}{0.5pt}}}1
}[keywords,comments,strings]
\newcommand{\sumlemma}{71,795}
\newcommand{\sumdef}{27,481}
\newcommand{\sumproject}{32}
\newcommand{\toolname}{\textsc{TheoremExtr}}

\begin{document}
\title{Extraction and Search in Rocq: Theorems, Definitions and Their dependencies}
%
%


\author{Jian Fang \and
Yingfei Xiong\textsuperscript{(\Letter)}
}
\authorrunning{J Fang \and Y Xiong}
%
\institute{Key Laboratory of High Confidence Software Technologies (Peking University), Ministry of Education; School of Computer Science, Peking University, Beijing, China
\email{fangjian@stu.pku.edu.cn, xiongyf@pku.edu.cn} 
}
\maketitle              
\begin{abstract}
Rocq (Coq) are now widely used in various fields, including software verification and mathematical proofs.
When proving a new theorem, users often need to search and apply proven theorems to assist the current proof process. 
However, the current search command is limited to the environment of imported modules and cannot search for theorems outside of this scope. 
Furthermore, tool developers and researchers may want to obtain detailed information about theorems, such as theorem's names, statements, and dependencies.
But there are currently no user-friendly and efficient tools available for extracting comprehensive information from Rocq projects.
We introduce a Rocq theorem extraction and analysis tool, \toolname{}, which is capable of analyzing theorem composition and extracting theorems, dependencies, and definitions from both parsing phase and runtime. 
We extracted \sumlemma{} theorems and their dependencies from \sumproject{} open-source projects from the Rocq community.
In addition, we extracted \sumdef{} definitions and their types among these projects.
We also developed a website that supports cross-project similarity search for theorems and definitions.
The tool is available at \url{https://github.com/Rw1nd/TheoremExtr}, and the search website is available at \url{https://lemmasearch.com/}.

\end{abstract}

\section{Introduction}
Interactive theorem provers (ITPs) such as Rocq~\cite{CoqRefMan8.20}, are widely employed in computer science~\cite{DBLP:conf/sosp/KleinEHACDEEKNSTW09,DBLP:conf/itp/KrebbersLW14} and mathematics~\cite{mahboubi2021mathematical}.
Users construct formal proofs by manually writing proof scripts and need to reference previously proven theorems.
However, Rocq currently lacks a unified theorem repository, like mathlib4 in Lean 4~\cite{mathlib4}.
Users have to manually search within projects to determine whether the  required theorems already exist.
Furthermore, researchers who wish to extract theorems from Rocq projects must acquire additional development knowledge (e.g., language server protocol), which increases the barrier.
These give the following challenges:

\paragraph{The search methods is limited.}
The \texttt{Search} command in Rocq is restricted to searching theorems within the current run-time environment.
It cannot perform cross-project searches or access theorems that have not been imported into the active environment.
This limitation imposes an additional burden on users, who must manually search through open-source projects to determine whether required theorems exist.

\begin{figure}[ht]
    \centering
    \includegraphics[width=0.8\linewidth]{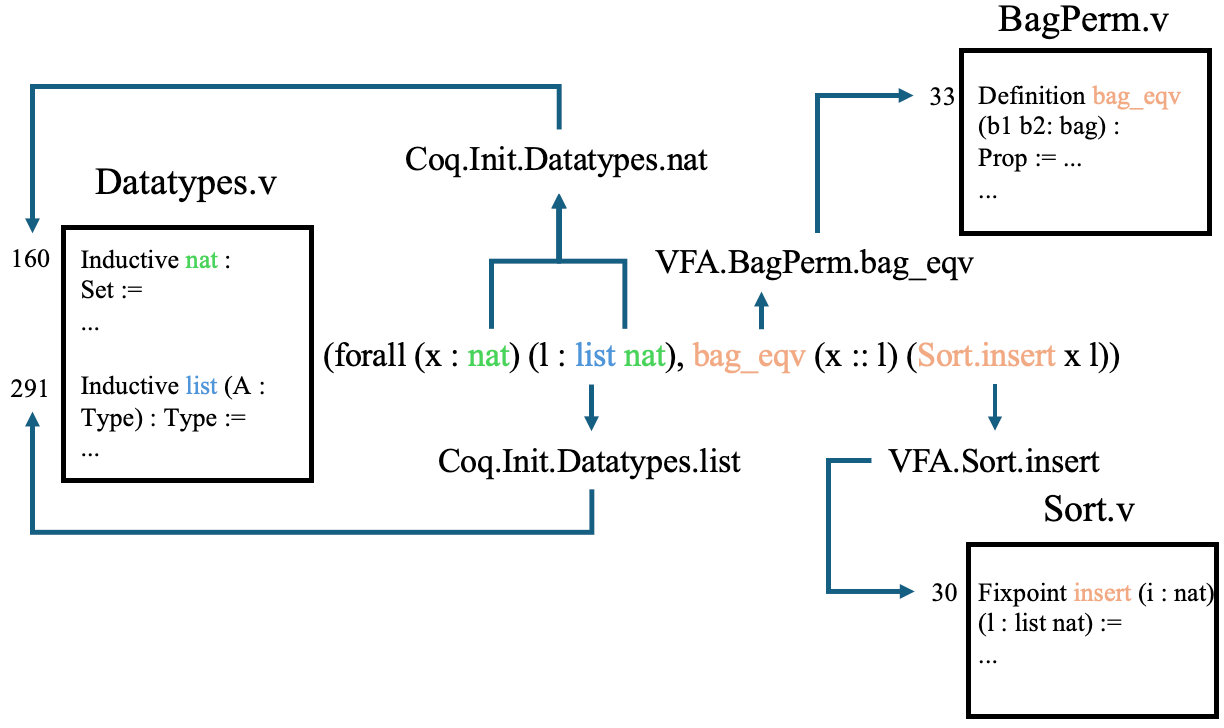}
    \caption{Theorem and its dependencies}
    \label{fig:lemma}
\end{figure}

\paragraph{Theorem extraction is challenging.}
For researchers in domains such as automated theorem proving and large language models (LLMs) for formal methods~\cite{DBLP:conf/icse/ThompsonSCFSB0L25,DBLP:conf/kbse/LuD024}, extracting theorems from Rocq projects is essential to obtain comprehensive information (e.g., types, dependent functions).
As illustrated in Figure \ref{fig:lemma}, the theorem demonstrates a property of insertion sort in \cite{Appel:SF3}.
This example shows that a single theorem may depend on numerous definitions distributed across multiple files.
However, mature and user-friendly tools to assist researchers in extracting those information are currently unavailable.

\textbf{Our work.}
To address these challenges, we developed \toolname{}, a theorem extraction and analysis tool.
The tool collects data from both Rocq's parsing phase and runtime to extract comprehensive information for each theorem, including its statement, dependent functions, and type definitions. 
Users can obtain detailed information from a project using a single command and a simple Python script, requiring minimal Rocq knowledge.
Notably, we use the runtime only to analyze theorem dependencies and type information, without extracting runtime execution state information, such as stack or heap data.
The contributions of this paper are as follows:
\begin{itemize}
    \item We implemented \toolname{}, a tool that extracts theorems and their internal dependency definitions, achieving a balance between run-time efficiency and user-friendliness.
    \item We extracted theorems from the Rocq platform, a distribution of the Rocq prover with libraries and plugins, as well as other open-source projects.
    Based on the extracted theorems and definitions, we developed a website available at \url{https://lemmasearch.com}.
    The website provides cross-project search capabilities with similarity search and localization to original project sources.
\end{itemize}

\section{Approach}
We observe that relying only on syntactic-level extraction in the parsing phase has significant limitations. 
For instance, a function definition within the current theorem might depend on files imported from external files.
Moreover, the definition of a theorem may contain implicit arguments, which need to be analyzed according to the specific runtime context.
And extracting information exclusively at runtime fails to capture precise scope information and line numbers for each theorem.

\begin{figure*}[ht]
    \centering
    \includegraphics[width=1\linewidth]{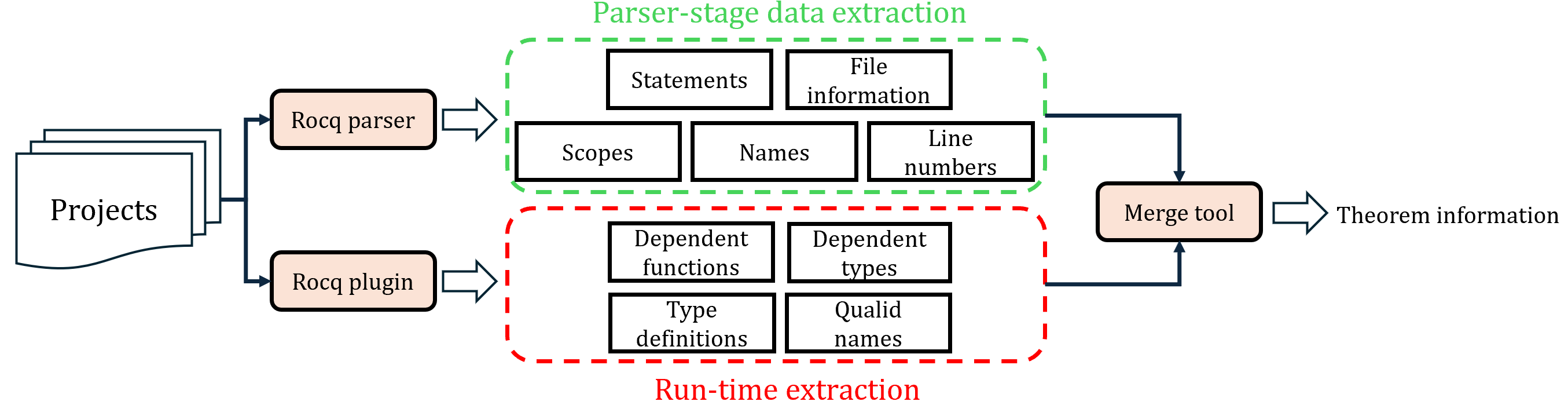}
    \caption{Overview of \toolname}
    \label{fig:overview}
\end{figure*}

Therefore, we need to combine the results of both analyses. 
Our tool \toolname{}, comprises three components, as shown in \Cref{fig:overview}: i) Parser-stage data extraction, which is integrated into the Rocq compiler and extracts theorem-related and file information.
ii) Run-time extraction that extracts data during runtime. 
iii) The merging tool that combines the parser-stage data and run-time information.
Our tool takes a Rocq project as input and outputs theorems-related data from the project.

\subsection{Parser-stage data extraction}
\label{sec:stage1}
This component extracts data from the syntax by modifying portions of the Rocq compiler code.
The component retrieves data from the following elements: i) statements (theorem statements or definition declarations), ii) scopes (modules, segments), iii) file information (file names, file path), iv) names (theorem names or definition names), v) line numbers.
To retrieve various types of data, we appropriately integrate our data extraction code based on the data structures defined in the Rocq source code.
The extraction methods for different elements are described below.

\textbf{Statements and names}.
Building upon the original logic of the Rocq compiler, we add auxiliary functions to extract theorem statements, definitions, and their names based on the data structure that represents the abstract syntax tree. 
The detailed definition of this data structure can be found in \cite{rocqapi}. 
And our code is implemented without altering the original compiler logic.

\toolname{} identifies each theorem by using its absolute path combined with the theorem name. 
When we use the nature number type \texttt{nat}, the name of \texttt{nat} is \texttt{Coq.Init.Datatypes.nat}.
To enable a one-to-one correspondence with theorems or definitions, we also record the name of the file being compiled and its absolute path.

\textbf{Scopes}.
To correlate the theorems within a module with the information obtained during run-time, we need to manually manage the scope of the module during the parser-stage extraction.
In Rocq, modules can contain theorems and proofs.
When the parser processes module-related code, we identify the data to be extracted by marking the current scope.
Theorems defined between the start and end of a module are considered internal to the module.

\textbf{File information and line numbers}.
When the Rocq compiler compiles a file, it locates the corresponding file based on its file path. 
During code parsing, the compiler records the location of each theorem and definition within the file. 
We capture this information during compilation for the files.

With these approaches, the data for each theorem or definition can be extracted at compile time. Our extraction method has a negligible impact on compilation efficiency.

\subsection{Run-time Extraction}
\label{sec:stage2}
Since Rocq supports extending its commands through plugins, we extract data from Rocq's environment by implementing a Rocq plugin.
Rocq plugins execute at run-time and can access all information within the current environment.

To extract theorem-related information, we first retrieve all theorems accessible in the current environment. 
Due to the syntactic omission in some theorem types, we need to analyze their concrete types to obtain complete type definitions.
And we analyze each subterm in the theorem to determine whether it is a function or an inductive type.
For the inductive types used in theorems, we extract their concrete definitions. 
To align with the data from the parser-stage data, the plugin must also record the absolute path and the name of each theorem.

\subsection{Merging Tool}
\label{sec:stage3}
The merging tool combines the data obtained from parser-stage extraction with the data extracted at run-time.
\toolname{} use the names of theorems to merge information for the same theorems or definitions.
For data obtained from parser-stage extraction, we reconstruct the complete full name by combining the compilation path, filename, scope, and theorem name.
For data obtained at run-time, the theorem names are already full names and can be used directly.
By merging the data from these two sources, \toolname{} ultimately provides complete information for each theorem or definition and saves it to a JSON file.

\section{Implementation and Usage}
This section introduce the implementation and usage of extraction tool \toolname{} and the search website.

\textbf{Extraction Tools}.
We implemented \toolname{} on Rocq 8.20.0.
We modified the codes in the Rocq compiler to extract data from the parse phase.
Use \texttt{coqc} command to compile the Rocq files, and it will generate a new JSON file in the target directory.
Additionally, we implemented a Rocq plugin that introduces a new command, \texttt{Createdb}, enabling the extraction of data from run-time. 
To use the plugin, users must first import all libraries to be extracted and then invoke the \texttt{Createdb} command, as shown in the following code:
\begin{lstlisting}[language=Coq, frame=single]
Require Import Target.Lib.
...
Createdb.
\end{lstlisting}
When the plugin completes extraction, a new JSON file is generated in the default path.
The Python script merges these two parts of data.

All experiments were conducted on a machine equipped with an Intel Core Ultra 7 265K processor and 48\,GB of RAM, running Ubuntu 22.04.5 LTS under the Windows Subsystem for Linux (WSL).
Table~\ref{tab:time} reports the time overhead per project, measured as the combined cost of parser-stage data extraction and runtime extraction.

\begin{longtable}{lrrr}
\caption{Time overhead per project (in seconds).}\label{tab:time}\\
\toprule
Project & \shortstack[l]{Parser-stage\\ data extraction (s)} & \shortstack[l]{Runtime\\ extraction (s)} & Total (s) \\
\midrule
\endfirsthead
\toprule
Project & \shortstack[l]{Parser-stage\\ data extraction (s)} & \shortstack[l]{Runtime\\ extraction (s)} & Total (s) \\
\midrule
\endhead
coq-aac-tactics & 4 & 5 & 9 \\
mathcomp-algebra-tactics & 12 & 29 & 41 \\
mathcomp-analysis & 447 & 157 & 604 \\
mathcomp-bigenough & 1 & 6 & 7 \\
coq-bignums & 11 & 10 & 21 \\
coq-itauto & 55 & 5 & 60 \\
compcert & 254 & 75 & 329 \\
coq-corn & 172 & 1679 & 1851 \\
coq-equations & 6 & 1 & 7 \\
coq-ext-lib & 4 & 7 & 11 \\
coq-fcsl-pcm & 41 & 51 & 92 \\
coq-gappa & 13 & 10 & 23 \\
coq-hott & 41 & 37 & 78 \\
coq-htt & 39 & 28 & 67 \\
coq-relation-algebra & 37 & 7 & 44 \\
coq-stdlib & 79 & 198 & 277 \\
coq-coqeal & 61 & 59 & 120 \\
coq-coqprime & 27 & 21 & 48 \\
coq-coquelicot & 34 & 10 & 44 \\
mathcomp-finmap & 10 & 7 & 17 \\
coq-flocq & 66 & 14 & 80 \\
coq-interval & 85 & 26 & 111 \\
coq-iris & 126 & 1369 & 1495 \\
coq-math-classes & 28 & 18 & 46 \\
mathcomp & 238 & 14 & 252 \\
coq-mtac2 & 18 & 6 & 24 \\
mathcomp-multinomials & 25 & 31 & 56 \\
coq-quickchick & 12 & 8 & 20 \\
mathcomp-real-closed & 47 & 34 & 81 \\
coq-reglang & 19 & 16 & 35 \\
coq-stdpp & 43 & 147 & 190 \\
coq-vst & 509 & 143 & 652 \\
\midrule
Total & 2564 & 4226 & 6790 \\
\bottomrule
\end{longtable}

\begin{figure*}[ht]
    \centering
    \includegraphics[width=1\linewidth]{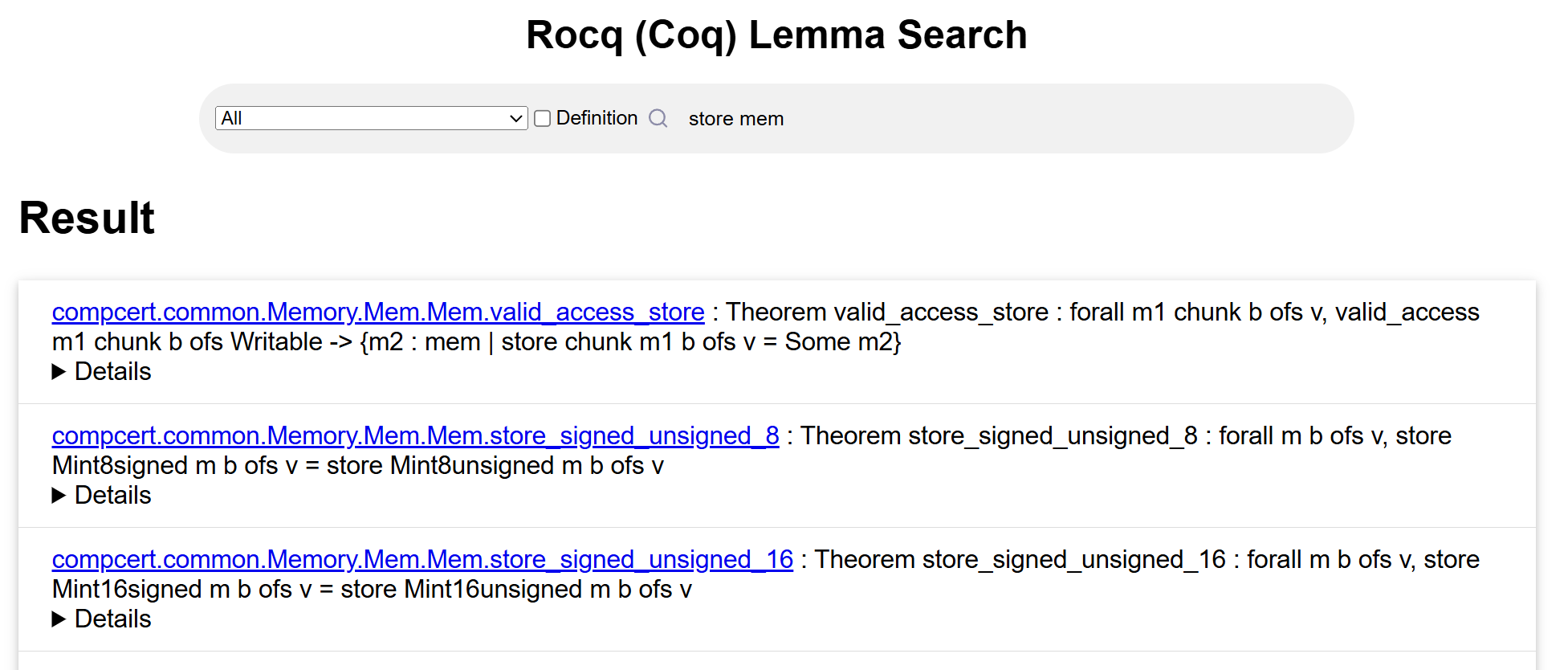}
    \caption{Theorem search website}
    \label{fig:website}
\end{figure*}
\textbf{Search website}.
We selected \sumproject{} open-source projects from the Rocq platform (version 2025.01.0) and the Rocq community for extraction, resulting in a total of \sumlemma{} theorems and \sumdef{} definitions.
We used Flask~\cite{grinberg2018flask} to build a theorem search website.
The website utilizes the BM25~\cite{bm25s} algorithm to enable similarity search.
As shown in Figure \ref{fig:website}, when searching for theorems related to memory storage, one can directly use \texttt{store mem} as the keywords for the search.
The search results show that the CompCert project contains relevant theorems.
In contrast, using a general search engine with the same keywords fails to find the related Rocq theorems.
Moreover, users need to know in advance which projects are relevant.
The results demonstrate that our website can search lemmas across projects and provides dependencies and associated types.
To enable the localization of theorems within their original projects, each theorem name is hyperlinked to its corresponding location in the original code repository.

\section{Related work}
Currently, several tools are available for extracting data from Rocq. CoqPyt~\cite{DBLP:conf/sigsoft/CarrottSTL0F24} is a Python-based extraction framework that relies on coq-lsp~\cite{coqlsp}. However, compared to \toolname{}, it lacks the capability to analyze dependency information within theorems.
Coq SerAPI~\cite{GallegoArias2016SerAPI} is another tool capable of extracting internal data from Coq. However, it requires users to learn additional protocol commands. In contrast, \toolname{} provides a more user-friendly interface, as it can be operated using simple Rocq commands.
Coq-lsp~\cite{coqlsp} can also be used for extracting data from Rocq. However, it requires additional knowledge of the LSP protocol~\cite{lsp}, which is not specific to Rocq.

And there are also serveral theorem datasets, such as Coqgym~\cite{yang2019learningprovetheoremsinteracting} and CoqStoq~\cite{DBLP:conf/icse/ThompsonSCFSB0L25}. However, these datasets are either outdated or lack scalability, making it challenging to migrate across different Rocq versions.
\toolname{} is implemented based on the Rocq compiler and its plugins, making it easily portable across different versions of Rocq.
\section{Conclusion and future work}
We developed the tool \toolname{}, which is capable of extracting theorems and definitions during the parsing and run-time phases of Rocq. \toolname{} is user-friendly and has a low learning curve.
Using \toolname{}, we extracted \sumlemma{} theorems and \sumdef{} definitions from \sumproject{} open-source projects and developed a search website to support similarity search.
Moreover, since our approach is built directly upon Rocq itself, it offers excellent scalability.

In addition, \toolname{} can be used in LLM training or within LLM agents.
\toolname{} can access richer theorem information, it provides more Rocq corpora for training LLMs and supplies more informative feedback for LLM agents.

Currently, \toolname{} has been implemented on Rocq version 8.20.0. Due to differences between Rocq versions, we plan to extend the implementation to all future versions of Rocq.

%
%
%
\bibliographystyle{splncs04}
\bibliography{mybib}

\end{document}